\documentclass[aip,jcp]{revtex4-2} % {traditional, twocolumn}
\usepackage[T1]{fontenc} % Use modern font encodings
\usepackage[version=3]{mhchem} % Formula subscripts using \ce{}
\usepackage[dvipsnames]{xcolor}
\usepackage{graphicx,subcaption}
\usepackage{amssymb}
\usepackage{overpic}
\usepackage{braket}
\usepackage{comment}
\usepackage{soul}
\usepackage{tablefootnote}
\usepackage{bm}
\usepackage{hyperref}
\hypersetup{pdfborderstyle={/S/U/W 0.5}}
\usepackage{booktabs}% http://ctan.org/pkg/booktabs
\usepackage{enumitem}
\usepackage{multirow}
\usepackage{tikz}
\usepackage{svg}
\usepackage{siunitx}

\usepackage{bibunits}
\defaultbibliographystyle{unsrt}

\newcommand{\hH}{{\hat{H}}}

\newcommand{\bz}{{\bf z}}
\newcommand{\bP}{{\bf P}}

\newcommand{\bd}{{\bf d}}
\newcommand{\br}{{\bf r}}
\newcommand{\bl}{{{\bf l}}}
\newcommand{\bp}{{\bf p}}
\newcommand{\bGamma}{{\bf \Gamma}}

\newcommand{\htheta}{{\hat{\theta}}}

\newcommand{\hbp}{{\hat{\bf p}}}

\newcommand{\hbr}{{\hat{\bf r}}}

\newcommand{\hbJ}{{\hat{\bf J}}}

\newcommand{\hbGamma}{{\hat{\bf \Gamma}}}

\newcommand{\bR}{{\bf R}}

\begin{document}

\author{Ben Curlee}
\affiliation{Department of Chemistry, Princeton University, Princeton, New Jersey 08544}

\author{Zheng Pei}
\affiliation{Chemistry Department, 
Brandeis University
%415 South Street
Waltham, MA 02453-2728}

\author{Xinchun Wu}
\affiliation{Department of Chemistry, Princeton University, Princeton, New Jersey 08544}

\author{Titouan Duston}
\affiliation{Department of Chemistry, Princeton University, Princeton, New Jersey 08544}

\author {Todd J. Martinez}
\affiliation{
Department of Chemistry, Stanford University, Palo Alto CA 94305
}

\author{Yihan Shao}
\affiliation{Chemistry Department, 
Brandeis University
Waltham, MA 02453-2728}

\author{Joseph E. Subotnik}
\email{subotnik@princeton.edu}
\affiliation{Department of Chemistry, Princeton University, Princeton, New Jersey 08544}

\title{Communication: Becke Weights as a Partitioning Scheme for Phase Space Electronic Structure Theory }

\begin{abstract}
We implement the Becke weight for atom A ($w_A$) as the partitioning scheme needed for phase space electronic structure theory. We show that this ansatz can lead to very efficient electronic structure calculations at nonzero nuclear momentum ($P \ne 0$). We further show that using Becke weights leads to efficient VCD calculations that are very stable with respect to basis set size. We hypothesize that, as far as quantum chemistry simulations in an atomic orbitals basis, the choice of Becke weights as a partitioning scheme will be adopted for all future phase space electronic structure calculations in an atomic orbital basis, as theorists learn to efficiently simulate coupled nuclear electronic dynamics through a new framework.
\end{abstract}

\maketitle

\section{Introduction}
Phase space electronic structure theory offers us the chance to explore non-Born Oppenheimer physics within a putatively reasonable  cost. The basis premise is that, if one is prepared to diagonalize an electronic structure Hamiltonian that depends on both nuclear position $\bR$ and nuclear momentum $\bP$, $H_{PS}(\bR,\bP)$, one can extract a great deal of electronic information that is not available within standard Born-Oppenheimer (BO) calculations, including electronic momentum\cite{nafie:1983:jcp:el_momentum,patchkovskii:2012:jcp:electronic_current}, electronic coriolis forces, and electronic centrifugal forces\cite{coraline:basisfree:2025}.  

For example, consider the case of electronic momentum.  When running nuclear dynamics on a Born-Oppenheimer ground state surface (with energy $E_0(\bR)$ and with electronic wavefunction $\Psi_0(\br;\bR)$), a naive calculation would predict that the electronic momentum is zero; after all, it is true that $\left<\Psi_0 \middle| \hat{\bp} \middle| \Psi_0 \right> = 0 $. Of course, this conclusion is incorrect because the nuclei are moving which yields its own contribution to the electronic momentum. As shown long ago by Nafie\cite{Nafie1983}, the correct result should be: 
\begin{eqnarray}
    \left<\hat{\bp}\right>  \stackrel{should \; be}{=} m_e \frac{d \left<\Psi_0 \middle| \hat{\br} \middle| \Psi_0 \right>}{dt}
\end{eqnarray}
\noindent where $\hbr$ is the electronic position operator, $m_e$ is the electron mass and $t$ is time. Of course, the resolution to the above paradox is to realize that the expectation value of the momentum must be taken relative to an electronic wavefunction that has been perturbed by nuclear motion. If the nuclear momentum $\bP$ is considered to be a classical object, the first order corrected wavefunction is a sum over all excited Born-Oppenheimer state ($j$):
\begin{eqnarray}
    \ket{\tilde{\Psi}_0} := \ket{\Psi_0} - i \hbar \sum_{j>0} \frac{\bP \cdot \bd_{0j}}{E_0 - E_j}\ket{\Psi_j}
\end{eqnarray}
Here, the matrix elements that couple the adiabatic electronic states to the nuclear momentum are the derivative couplings, $\bd_{jk} = \left< \psi_j\middle | \boldsymbol{\nabla} \psi_k \right>$, that are the first order corrections to Born-Oppenheimer theory.
A simple calculation\cite{coraline:2024:jcp:pssh_conserve} then demonstrates that indeed
\begin{eqnarray}
\left<\tilde{\Psi}_0 \middle| \hat{\bp} \middle| \tilde{\Psi}_0 \right>
= m_e \frac{d \left<\Psi_0 \middle| \hat{\br} \middle| \Psi_0 \right>}{dt} \label{eq:nafie}
\end{eqnarray}
which can also be written as:
\begin{align}
 \frac{\partial \left< \hbr \right>}{\partial \bR} =  \frac{m_e}{M}
  \frac{\partial \left< \hbp \right>}{\partial \bP} 
    \label{eq:nafie2}
\end{align}
\noindent Here the expectation value of $\bp$ is calculated with $\tilde{\Psi}_0$ and the expectation value of $\br$ is calculated with $\Psi_0$ (or $\tilde{\Psi}_0$). The equality in Eq. \ref{eq:nafie} demonstrates the power of  phase space electronic structure theory. Whereas $\ket{\Psi_0}$ is a function of nuclear position $\bR$, $\Psi_0(\br;\bR)$, $\ket{\tilde{\Psi}_0}$ is a function of both nuclear position $\bR$ and nuclear momentum $\bP$, $\tilde{\Psi}_0(\br;\bR,\bP)$.
To that end, over the last several years, our research teams have now derived meaningful phase space electronic structure Hamiltonians by approximating the derivative couplings by a one-electron operator $\hbGamma$ whose form is known {\em a priori}. Thus far, our best ansatz for $\hbGamma$ has been motivated by our desire to recapitulate the relevant symmetries of the derivative couplings --  in particular, the translational and rotational properties of the derivative coupling. These constraints lead to  a phase space electronic structure Hamiltonian of the form:

\begin{align}
    &\hat{H}_{PS}(\bR,\bP) =  \sum_I \frac{\left(\bP-i\hbar\hbGamma_I(\bR)\right)^2}{2M_I}+\hH_{el}(\bR)\label{PS_ham}\\
    &\hbGamma_A = \hbGamma_A'+\hbGamma_A''+\hbGamma_A'''
\end{align}
Here $M_I$ is the mass of nucleus $I$ and $\hH_{el}(\bR)$ is the electronic Hamiltonian at nuclear position \bR.

Above, $\hbGamma'$ enforces the correct translational symmetry (and therefore conservation of linear momentum) and $\hbGamma''$ enforces the correct rotation symmetry (and therefore conservation of angular momentum).  For systems where spin-orbit coupling is important (and the total angular momentum has a non-trivial spin component), one can also include  $\hbGamma'''$  which deals with the rotational properties of the spin component. Note that, although we have framed the discussion above in terms of recovering symmetry properties of the derivative couplings, one can also view these different $\hbGamma$ terms as being corrections to the electronic Hamiltonian that allow us to solve for the electronic steady states in terms of a moving, non-inertial nuclear frame. For instance, $\hbGamma''$ includes the electronic coriolis force that arises when we align the electronic frame with the nuclear frame.
For the exact mathematical equations, see Sec. \ref{theory} below.

In our earliest phase space electronic structure papers\cite{coraline:2024:jcp:pssh_conserve, tian:2024:jcp:erf,duston:2024:jctc_vcd}, we constructed $\hbGamma$ differently for every basis set. A breakthrough came when we realized that one need not invoke any given basis set (and avoid the need for any new phases on the atomic orbitals\cite{nafie:1992:vcd,Ditler:2022:NVP_VCD}). For instance, to  account for the translation  of the nuclei, one can set set:
    \begin{align}
    &\hbGamma_A' = \frac{1}{2i\hbar}\left(\hat{\theta}_A\hbp+\hbp\hat{\theta}_A\right)\label{gamma_start}
    \end{align}
Here (and below), the function $\htheta_A$  is  central to phase space (PS) electronic structure theory.  This function ($\htheta_A(\br)$) is a 
 partition of space function
 that is  $(i)$ always positive, $(ii)$ near unity when $\br$ is close to $\bR_A$, $(ii)$ zero when $\br$ is far from $\bR_A$, and $(iv)$ always satisfies $\sum_A\hat{\theta}_A=1$.
 Physically speaking, the function $\htheta_A$ allows us to partition electronic density by atom, so that electronic density near atom $A$ will be boosted by the momentum of atom $A$. 
 Thus, the form of $\htheta_A$ is of utmost importance.

Thus far, in all of our group's basis-free publications in this area\cite{coraline:basisfree:2025,xuezhi:cpr:review:2026}, we have always used a simple partition scheme known as Hirshfeld partitioning ($\theta_A = w_A^{Hirsh}$):
\begin{eqnarray}
    w_A^{Hirsh}(\hbr) = \frac{Z_Ae^{(\hbr-\bR_A)^2/\sigma_A^2}}{\sum_B Z_Be^{(\hbr-\bR_B)^2/\sigma_B^2}}\label{hirshfeld_def}
\end{eqnarray}
Here $Z_A$ is the nuclear charge of nucleus $A$ and $\sigma_A$ is a free parameter that defines the length scale. That being said, the form in Eq. \eqref{hirshfeld_def} has several shortcomings. First, $\htheta_A$ may be spread over many atoms and so will not allow to easily understand electron transfer between two atoms.  Second, $\htheta_A$ can also have a sharp cutoff between atoms and become numerically difficult.  Third, the nature of the Hirshfeld partitioning really depends critically on ratios between the $\sigma_B$ and the internuclear distances. In particular, for a given $\sigma$ in Eq. \eqref{hirshfeld_def}, the shape of the partition in a diatomic molecular  depends on the internuclear distance. Strangely, the cutoff gets sharper as the atoms get farther apart.

Within this lens, a seasoned quantum chemist will of course note that there is an alternative to Hirshfeld partitioning. Namely, we can use Becke weights.\cite{becke:1988:jcp:beckegrid,frisch:cpl:1996:beckegrid,toddmartinez:book:beckegrid} As a reminder, Becke weights arise when one wants to evaluate a function $f$ in three dimensional space by partitioning over atomic grids. Namely, one builds a set of normalized functions $w_A$ satisfying 
\begin{align}
    \sum_A w_A^{Becke}(\br) = 1
\end{align}
and so the integral becomes 
\begin{align}
I &= \int dr f(r) = \int dr \sum_A f(r) w_A(r) \\
 &\approx  \sum_A\sum_{G_A} f(G_A) w_A(G_A) W(G_A)\label{integral_approx}
\end{align}
where $W(G_A)$ is the Lebedev weight for integration on atom A's grid.
Becke partitioning is less sensitive to any potential parameters than is Hirshfeld partitioning in the sense that, for Becke, one also knows 
that $w_A^{Becke}(\br = \bR_A) = 1$ and $w_A^{Becke}(\br = \bR_{B\ne A}) = 0$; 
these constraints need not hold for Hirshfeld partitioning.  
Becke weights can also be constructed efficiently nowadays, and thus Eq. \eqref{integral_approx} have been used very fruitfully over the years to evaluate density functional theory (DFT) functionals. Moreover, it is also well known today that Becke partitioning can offer  an improved set of atomic charges\cite{Mei:2015:partitioning_polarizablility} (relative to Hirshfield) such that: 
\begin{align}
\langle\sum_A Q_A\bR_A\rangle\approx\langle\hat{\mu}\rangle  \\
Q_A \equiv Z_A-\int\theta_A(\br)\rho(\br) d\br
\end{align}
Here $\rho(\br)$ is the charge density at position $\br$.

The discussion above raises the simple question: why not set \begin{align}
    \theta_A(\br) = w_A^{Becke}(\br)
\end{align}
so as to allow PS calculations to  proceed both  accurately (using the improved atomic charges from Becke weights) and efficiently  (using modern electronic structure grid subroutines).

We have now implemented the ansatz above within a developmental version of Q-Chem\cite{qchem6}. As far as validating the ansatz,
the natural calculation to run is vibrational circular dichroism (VCD). As a reminder, VCD arises because chiral enantiomers absorb slightly different amounts of left and right handed light, and the approach can be used to help determine the structure of natural products, most notably assisting with the absolute configuration\cite{Polavarapu:2020:VOA_structure_determination}.  When averaged over different incoming light directions, this difference is known as the rotary strength, and takes the form $\mathcal{R} = \operatorname{Im}(\bm{\mu}_{if}\cdot\textbf{m}_{fi})$ where $\bm{\mu}_{if}$ is the electronic transition matrix element and $\textbf{m}_{fi}$ is the magnetic transition matrix element (which is not readily accessible between vibrational states within BO theory).
In particular, to calculate the electronic component of ${\bf m}$, BO theory faces a severe problem in so far as the electronic magnetic moment (like the electronic momentum) is always zero within BO theory -- even when nuclei are moving. This problem was realized long ago by Nafie and Stephens and others (leading to a host of methods, including  nuclear velocity perturbation\cite{nafie:1992:vcd,Ditler:2022:NVP_VCD}, second order Møller–Plesset perturbation based approaches\cite{Shumberger:2025:MP2_VCD}, and the origin-invariant length gauge approach\cite{Shumberger:2026:OILG_VCD}). Nowadays, the most common resolution to this conundrum is to use what is known as magnetic field perturbation theory\cite{stephens:1985:jpcc_vcd} (MFP), whereby one expresses the dependence of ${\bf m}$ on nuclear momentum to be\cite{stephens:1985:jpcc_vcd}: 
\begin{align}
    \frac{\partial}{\partial P_{A\alpha}}\bra{\Psi_0}\hat{\textbf{m}}^e\ket{\Psi_0}_{eq} = \frac{2i\hbar}{M_A}\left\langle{\frac{\partial\Psi_0}{\partial\textbf{B}}}\middle |{\frac{\partial \Psi_0}{\partial R_{A\alpha}}}\right\rangle_{eq}
\end{align}
The expression above is effectively a berry-curvature result that makes clear that VCD can act as a sensitive electronic probe of nuclear momentum.

In this communication, we will use VCD to test our approximation to $\hbGamma$ in Equations \ref{gamma_start} and \ref{gamma_start2}.
Note that a phase space electronic structure approach automatically (unlike BO theory) recovers a nonzero VCD signal. Moreover, if one replaces $\hbGamma$ in Eq. \ref{PS_ham} with the full derivative coupling (which was the original PS aproach pioneered by Shenvi\cite{shenvi:2009:jcp_pssh}) then one does recover the exact MFP signal. That being said, the goal here is to use VCD spectra to learn about how to best approximate that derivative coupling and form the optimal $\hbGamma$.
A recent paper\cite{zhentao:2024:jcp:vcd_basis_free} has proven that Hirshfeld partitioning is able to recover VCD signals for the oxirane molecule, and so one would like to retain the accuracy found in Ref. \citenum{zhentao:2024:jcp:vcd_basis_free}. At the same time, it also true that previous implementations of electronic phase space theory have been limited because of slow computational speed. Here, we will show that, by  using well-benchmarked Becke weights within the Q-Chem package, we can achieve a very fast implementation of PS electronics structure theory, which we will compare against MFP.

\section{Theory}\label{theory}

In this communication, we will diagonalize a phase space electronic structure Hamiltonian with  $\hbGamma$ operators given in Eq. \ref{gamma_start} above and Eqs. \ref{gamma_start2}-\ref{gamma_end} below (from Ref. \citenum{xuezhi:cpr:review:2026}).
    To account for the rotation  of the nuclei, we set:
\begin{align}
    &\hbGamma_A'' = -\sum_B \zeta_{AB} \left(\bR_A - \bR_B^0\right)\times \left(\textbf{K}^{-1}_B\hbJ_B\right) 
    \label{gamma_start2}\\
    &\hbJ_B = \frac{1}{2i\hbar}\left(\left(\hbr-\bR_B\right)\times\left(\hat{\theta}_B\hbp+\hbp \hat{\theta}_B\right)\right)\\
    & \bR_B^0 = \frac{\sum_A \zeta_{AB}\bR_A}{\sum_A\zeta_{AB}}\\
    &\textbf{K}_B = \sum_A \zeta_{AB}\left(\left(\bR_A^\top\bR_A-\bR_B^{0\top}\bR_B^{0}\right)I_3-\left(\bR_A\bR_A^\top-\bR_B^0\bR_B^{0\top}\right)\right)\\
    &\zeta_{AB} = M_A e^{-(\bR_A-\bR_B)^2/\beta_{AB}^2}\label{gamma_end}
\end{align}

One can run electronic structure calculations for any $\bR$ and $\bP$ coordinates using the Hamiltonian above. 
In practice, we compute the key matrix element for Eq. \ref{gamma_start} as
\begin{align}
    (\theta_A p_x)_{\mu \nu} = \sum_{G_A} W(G_A) w_A(G_A)\phi_{\mu}(G_A)\phi_\nu^x(G_A)
\end{align}
where (following the usual notation) $W(G_A)$ is the Lebedev weight, $w_A(G_A)$ is the Becke weight, $\phi_\mu$ is an atomic orbital basis function, and $\phi_\nu^x$ is the derivative of basis function $\phi_\nu$ in the $x$ direction. In the above equation (and below), we have dropped the superscript ``Becke'' from $w_A$.

Finally, to calculate $\mathcal{R}$, note that for vibrational mode $k$  with displacement $S_{A \alpha k}$ for atom $A$ in direction $\alpha$, $\mathcal{R}$ takes the form\cite{duston:2024:jctc_vcd}: 
\begin{align}
    \mathcal{R}_k=\sum_{A\alpha A'\alpha'}\frac{\hbar M_A}{2}S_{A\alpha k}S_{A'\alpha' k}\left(\frac{\partial \textbf{m}_0}{\partial P_{A\alpha}}\right)_{\operatorname{eq}}\left(\frac{\partial \bm{\mu}_0}{\partial R_{A'\alpha'}}\right)_{\operatorname{eq}}
\end{align}
Here $\textbf{m}_0$ is $\bra{\Psi_0^{PS}}\hat{\textbf{m}}\ket{\Psi_0^{PS}}$, $\bm{\mu}_0$ is $\bra{\Psi_0^{PS}}\hat{\bm{\mu}}\ket{\Psi_0^{PS}}$, and $\Psi_0^{PS}$ is the ground state wavefunction of the PS Hamiltonian, $H_{PS}$. The operator $\hat{\bf m}$ is the magnetic moment, which is defined for an electron to be $\hat{\bf m} = -\frac{e}{2m_e} \hat{\bf l}$ (where $\hat{\bf l}$ is the electronic angular momentum). We calculate $\partial \bm\mu_0/ \partial R$ and  the hessian within BO theory (rather than PS theory) as these quantities are expected to be very similar within both BO and PS frameworks.
We ignore the  $\hbGamma^2$ term. (This term is typically very small.)

Following Ref. \citenum{duston:2024:jctc_vcd}, we will evaluate $\partial \textbf{m}/\partial P_{A \alpha}$
using a distributed origin (DO)\cite{stephens:1987:gauge:vcd} and  Nafie's relationship (Eq. \ref{eq:nafie2}) as follows:

\begin{align}
\left(\frac{\partial \textbf{m}_0}{\partial P_{A\alpha}}\right) &= \frac{-e}{2m_e} \frac{\partial }{\partial P_{A \alpha}} \left<\Psi^{PS}_0 \middle | \hbr \times \hbp \middle| \Psi^{PS}_0 \right> +   \frac{\partial }{\partial P_{A \alpha}} \sum_B \frac{Z_B}{2 M_B}\bR_B \times \bP_B \\
 &= \frac{-e}{2m_e}\frac{\partial }{\partial P_{A \alpha}} \left<\Psi^{PS}_0 \middle | (\hbr - \bR_A) \times \hbp \middle| \Psi^{PS}_0 \right>
 +
 \frac{-e}{2m_e}\bR_A \times
 \frac{\partial }{\partial P_{A \alpha}} \left<\Psi^{PS}_0 \middle |   \hbp \middle| \Psi^{PS}_0 \right>
 +
{\bf f}_\alpha(\bR_A)
 \\
 &= \underbrace{\frac{-e}{2m_e}\frac{\partial }{\partial P_{A \alpha}} \left<\Psi^{PS}_0 \middle | (\hbr - \bR_A) \times \hbp \middle| \Psi^{PS}_0 \right>}
 +
 \underbrace{\frac{-e}{2M_A} \bR_A \times
 \frac{\partial }{\partial R_{A \alpha}} \left<\Psi^{PS}_0 \middle |   \hbr \middle| \Psi^{PS}_0 \right> }
  +
{\bf f}_\alpha(\bR_A)
 \\
 & \equiv 
 \hspace{60pt}\left(\frac{\partial \textbf{m}_0^{DO}}{\partial P_{A\alpha}}\right) \hspace{50pt}+ 
\hspace{50pt}\left(\zeta_{
Stephens}^{L\cdot P}\right)  \hspace{15pt}+ 
\hspace{15pt}{\bf f}_\alpha(\bR_A)
\end{align}
Here, we have defined the vector function $f_{\alpha} (\bR_A)  = \sum_{\beta\gamma} \frac{Z_A}{2M_A}\epsilon_{\beta\alpha\gamma} R_{A\beta}\textbf{e}_\gamma$ where $\epsilon$ is the Levi-Civita symbol and $\textbf{e}_\gamma$ is the unit vector in the $\gamma$ direction. We have also identified $\zeta_{Stephens}^{L\cdot P}$ that is necessary for gauge invariance\cite{stephens:1987:gauge:vcd}. 

Now, to calculate 
$\left(\partial \textbf{m}_G/\partial P_{A\alpha}\right)$ within HF theory for an atomic orbital basis, let us write the atomic orbital (AO) one electron density matrix as $D_{\mu \nu} = \sum_i C_{\mu i } C_{\nu i}^*$ for molecule orbitals $C_{\mu i}$, so that:
\begin{align}
\left(\frac{\partial \textbf{m}_0^{DO}}{\partial P_{A\alpha}}\right) &= \frac{\partial}{\partial P_{A\alpha}}
\left( \sum_{\mu \nu} D_{\mu \nu} \textbf{m}^{DO}_{\mu \nu} \right)
\end{align}

If we now parametrize the molecular orbitals as\cite{maurice:thesis},
\begin{align}
    C_{\mu q} = \sum_i C_{\mu p}^0  \left(\exp(\kappa)\right)_{pq}
\end{align}
it follows that:
\begin{align}
\left(\frac{\partial 
\textbf{m}_0^{DO}}{\partial P_{A\alpha}}\right)
&=\frac{-e}{2m_e}\sum_{\mu\nu bj}C_{\mu b}\frac{\partial\kappa_{bj}}{\partial P_{A\alpha}}C_{\nu j}^*\left<\phi_\mu\middle | (\hbr-\bR_A)\times \hbp\middle | \phi_\nu\right>\\
&=\frac{-e}{2m_e}\sum_{\mu\nu bj}\frac{\partial\kappa_{bj}}{\partial P_{A\alpha}}\left<\psi_b\middle | (\hbr-\bR_A)\times \hbp\middle | \psi_j\right>\\
&=\frac{-e}{2m_e}\sum_{\mu\nu bj}\frac{\partial\kappa_{bj}}{\partial P_{A\alpha}}
\left( \bl_{bj} - \bR_A \times \bp_{bj} \right)\label{simple_cpscf}
\end{align}
Here $\psi_j$ and $\psi_b$ are occupied and virtual  molecular orbitals respectively. Finally, evaluating the response of the molecular orbitals to a perturbation in $\bP$ is particular easy (because no energy or overlap matrix elements depend on $\bP$) and requires merely a simple inversion
\begin{align}
    \sum_{ai} \left( \frac{\partial^2 E_{HF} }{\partial \kappa_{bj} \partial \kappa_{ai}} \right)
    \left( \frac{\partial \kappa_{ai} }{\partial  P_{A \alpha}} \right) = - \left( \frac{\partial^2 E_{HF}}{\partial \kappa_{bj} \partial  P_{A \alpha}} \right) = \frac{ i \hbar \Gamma^{A \alpha}_{bj}}{M_A}
\end{align}
where $E_{HF}$ is the Hartree-Fock energy. For efficiency, we use the standard z-vector trick and evaluate:
\begin{align}
    \left(\bz_l\right) _{bj} = \sum_{ai} \left( \frac{\partial^2 E_{HF} }{\partial \kappa \partial \kappa} \right)^{-1}_{bj, ai} \bl_{ai} \\
    \left(\bz_p\right) _{bj} = \sum_{ai} \left( \frac{\partial^2 E_{HF} }{\partial \kappa \partial \kappa} \right)^{-1}_{bj, ai} \bp_{ai}  
\end{align}
Note that we require only 6 (instead of $3N$ inversions) of Eq. \ref{simple_cpscf}. 
The final result is:
\begin{align}
\left(\frac{\partial 
\textbf{m}_0^{DO}}{\partial P_{A\alpha}}\right)
&=\frac{-i\hbar e}{2m_e}\sum_{\mu\nu bj}C_{\mu b}^*\bGamma^{A\alpha}_{\mu\nu}C_{\nu j}\left(\textbf{z}_l-\bR_A\times \textbf{z}_p\right)_{bj}\label{final_mDO}
\end{align}

The calculation of $\left(\frac{\partial \bm{\mu}_0}{\partial R_{A\alpha}}\right)_{\operatorname{eq}}$ is the same as within MFP, but a brief overview is given here for completeness. 
\begin{align}
    \left(\frac{\partial \bm{\mu}_0}{\partial R_{A\alpha}}\right)_{\operatorname{eq}} &=-e\frac{\partial }{\partial R_{A \alpha}} \left<\Psi_0 \middle | \br \middle| \Psi_0 \right> +  \frac{\partial }{\partial R_{A \alpha}} \sum_B Z_B\bR_B \\ 
    &=-e\frac{\partial }{\partial R_{A \alpha}}\left(\sum_{\mu\nu} D_{\mu\nu} \br_{\mu\nu}\right) + Z_A\textbf{e}_\alpha\\
    &=-e\sum_{\mu\nu} \left(\frac{\partial D_{\mu\nu}}{\partial R_{A\alpha}} \br_{\mu\nu}+D_{\mu\nu}\frac{\partial \br_{\mu\nu}}{\partial R_{A\alpha}}\right) + Z_A\textbf{e}_\alpha
\end{align}

\section{Results}

The theory above was incorporated into a developmental version of the Q-chem electronic structure package\cite{qchem6}. Before presenting VCD results, we note that as far as speed is concerned, the cost per self-consistent field (SCF) cycle of generalized Hartree-Fock (GHF) with and without PS (for oxirane in the cc-pvdz basis) is 0.143s and 0.132s.  Thus, future dynamical simulations with HF or DFT should not be hindered by incorporating PS corrections.
Note that, in this communication (except for one figure in the Appendix), we perform exclusively Hartree-Fock calculations. As discussed below, the $\hbGamma$ function is evaluated on a grid, but we emphasize that the grid is not used for any other purpose when computing VCD spectra -- e.g. we do not evaluate any DFT exchange-correlation energies on a grid.  This choice has been made on purpose so that one can clearly evaluate the change in accuracy and performance that arises when including $\hbGamma$ on a grid within a PS electronic structure calculation with (without any confusion as to the impact of the grid from any possible DFT artifacts).

\begin{figure}[ht]
    \centering
    \includegraphics[width=6.5in]{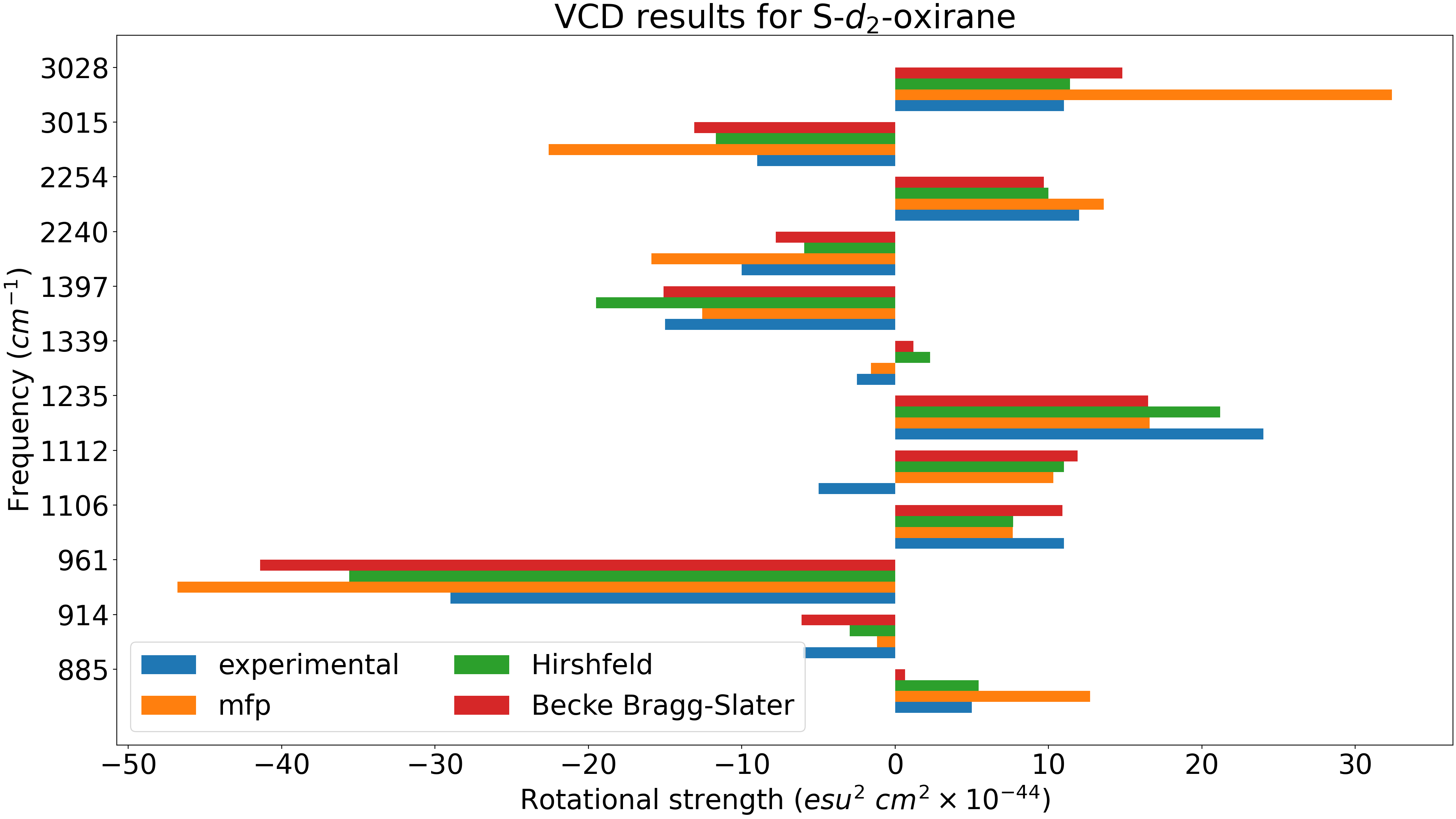}
    \caption{VCD results for three different methods compared with experiment for the molecule S-$d_2$-oxirane in a cc-pvdz basis. MFP indicates  magnetic field perturbation\cite{stephens:1985:jpcc_vcd}; Hirshfeld partitioning implements a PS approach using Eq. \ref{hirshfeld_def}, and Becke Bragg-Slater uses Eq. \ref{becke_function} with the Becke partition shifted using the Bragg-Slater radii. All calculations were performed at the restricted Hartree-Fock (RHF) level of theory.  }
    \label{VCD1}
\end{figure}

Let us now turn to VCD. 
Our key results are shown in Fig. \ref{VCD1} and Table \ref{dz_table}.
\begin{table}[!h]
\centering
\begin{tabular}{|llccrr|}
\hline
Frequency (cm$^{-1}$) & Experimental & MFP & Hirshfeld & Becke Uniform & Becke Bragg-Slater \\
\hline
672 & N/A & 1.03 & 1.67 & 2.04 & 1.57 \\
\hline
754 & N/A & 9.76 & 2.96 & 19.9 & 12.8\\
\hline
817 & (+) & 1.61 & 5.39 & 7.56 & 4.07 \\
\hline
885 & (5) & 12.7 & 5.45 & 15.6 & 0.638\\
\hline
914 & -6 & -1.18 & -2.96 & -17.5 & -6.11 \\
\hline
961 & -29 & -46.8 & -35.6 & -96.7 & -41.4 \\
\hline
1106 & 11 & 7.66 & 7.69 & 23.2 & 10.9 \\
\hline
1112 & -5 & 10.3 & 11.0 & 24.9 & 11.9 \\
\hline
1235 & 24 & 16.6 & 21.2 & 44.5 & 16.5 \\
\hline
1339 & (-2.5) & -1.58 & 2.27 & 3.09 & 1.18 \\
\hline
1397 & (-15) & -12.58 & -19.5 & -36.5 & -15.1 \\
\hline
2240 & -10 & -15.9 & -5.94 & -8.87 & -7.79 \\
\hline
2254 & 12 & 13.6 & 10.0 & 13.4 & 9.70 \\
\hline
3015 & -9 & -22.6 & -11.7 & -17.9 & -13.1\\
\hline
3028 & 11 & 32.4 & 11.4 & 16.2 & 14.8 \\
\hline
\end{tabular}
\caption{Experimental VCD spectrum\cite{Freedman:1991:vcd_oxirane_experiment} and predictions for S-$d_2$-oxirane.
Frequencies are from experiment. The MFP column is magnetic field perturbation. The Hirshfeld column is the Hirshfeld partitioning; Becke uniform is a Becke weight partitioning with equal weights to all types of nuclei; Becke Bragg-Slater is a Becke weight partitioning that gives a higher weight to atoms with higher estimated atomic radii. All calculations were performed using the cc-pvdz basis. Parentheses on experimental values indicate that they were estimated from relative intensities in the gas phase instead of integrated intensities in solution.\cite{Freedman:1991:vcd_oxirane_experiment} Experimental and theoretical results are in esu$^2$ cm$^2 \times 10^{-44}$.}
\label{dz_table}
\end{table}
In Fig. \ref{VCD1}, we plot experimental data\cite{Freedman:1991:vcd_oxirane_experiment}  for the S-$d_2$-oxirane molecule along with computer calculations using three different methods: MFP, PS with a Hirshfeld partitioning ($\sigma=1\textrm{\AA}$ for all atoms), and PS with a Becke weight partitioning shifted by the Bragg-Slater radii\cite{slater:1964:bragg-slater_radii}. The value of $\beta_{AB}$ was chosen to be 9.00 bohr for all atoms for all PS methods. The performance of all three methods is similar with the most notable difference being that only MFP gets the correct sign for the weak 1339cm$^{-1}$ signal. Note that a previous phase space implementation \cite{zhentao:2024:jcp:vcd_basis_free} recovered the correct  sign using the aug-cc-pvQZ  geometry and modes; the mode signal is very weak and the sign of the result is clearly very sensitive to the choice of basis and the form of the partitioning function. As has been found previously, all calculations fail for the 1112 cm$^{-1}$ signal (which is due to a lack of electron-electron correlation\cite{stephens:1994:cpl:exp_vcd_cycp}, which can strongly change the $\partial \mu/\partial \bR$ term).

\begin{figure}[ht]
    \centering
    \includegraphics[width=6.5in]{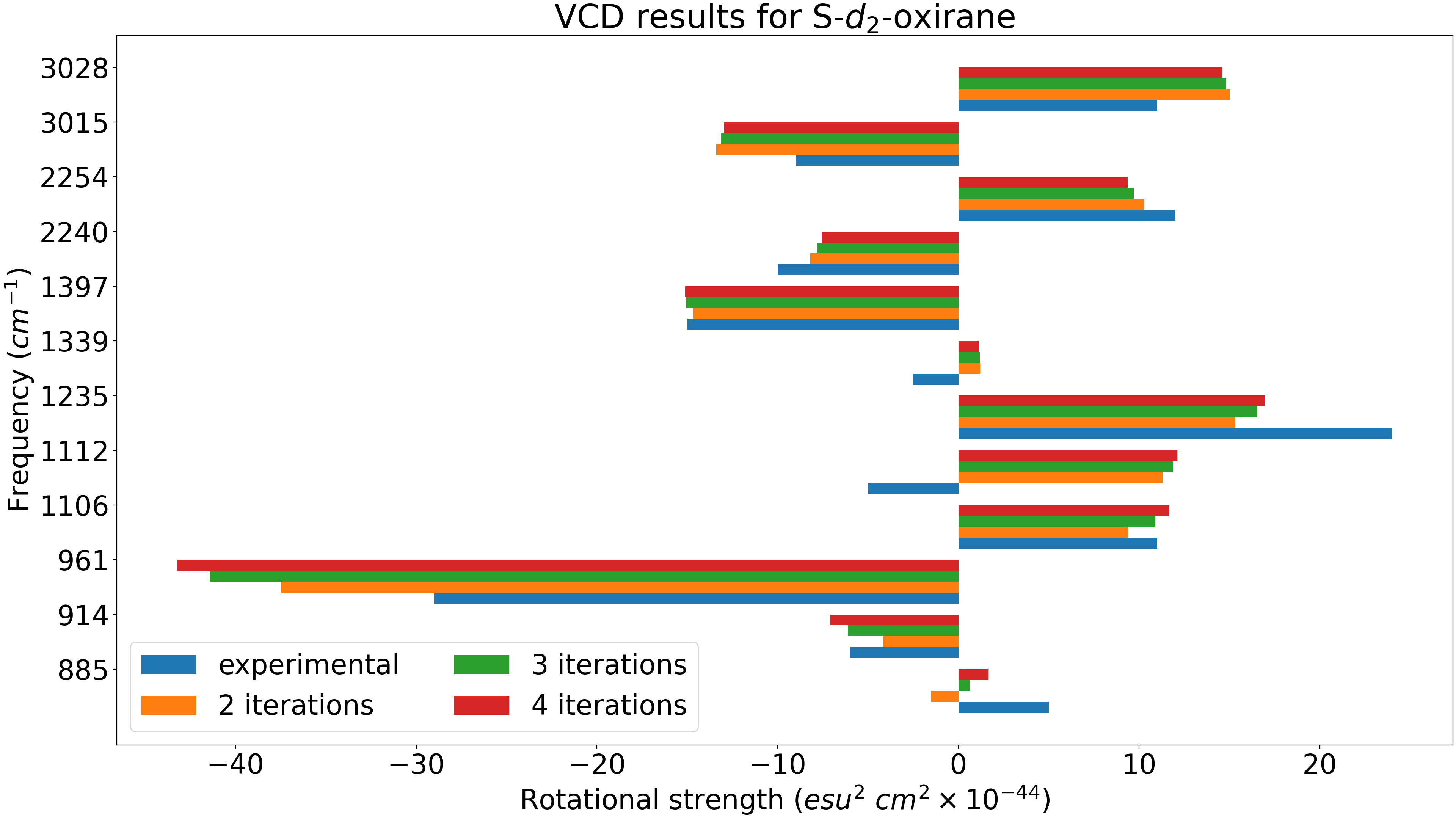}
    \caption{VCD results for the Becke partitioning shifted by the Bragg-Slater Radii with different numbers of iterations for the function in Eq. \ref{becke_function} that dictate the sharpness of the transition between atomic domains.  Note that the results are fairly insensitive to the number of iterations, indicating that using Becke weights is quite robust. The mean absolute error (relative to experiment) is 4.95 for 2 iterations, 4.80 for 3 iterations, and 4.99 for 4 iterations. Errors are in esu$^2$ cm$^2 \times 10^{-44}$.
    }
    \label{VCD3}
\end{figure}

\begin{figure}[ht]
    \centering
    \includegraphics[width=6.5in]{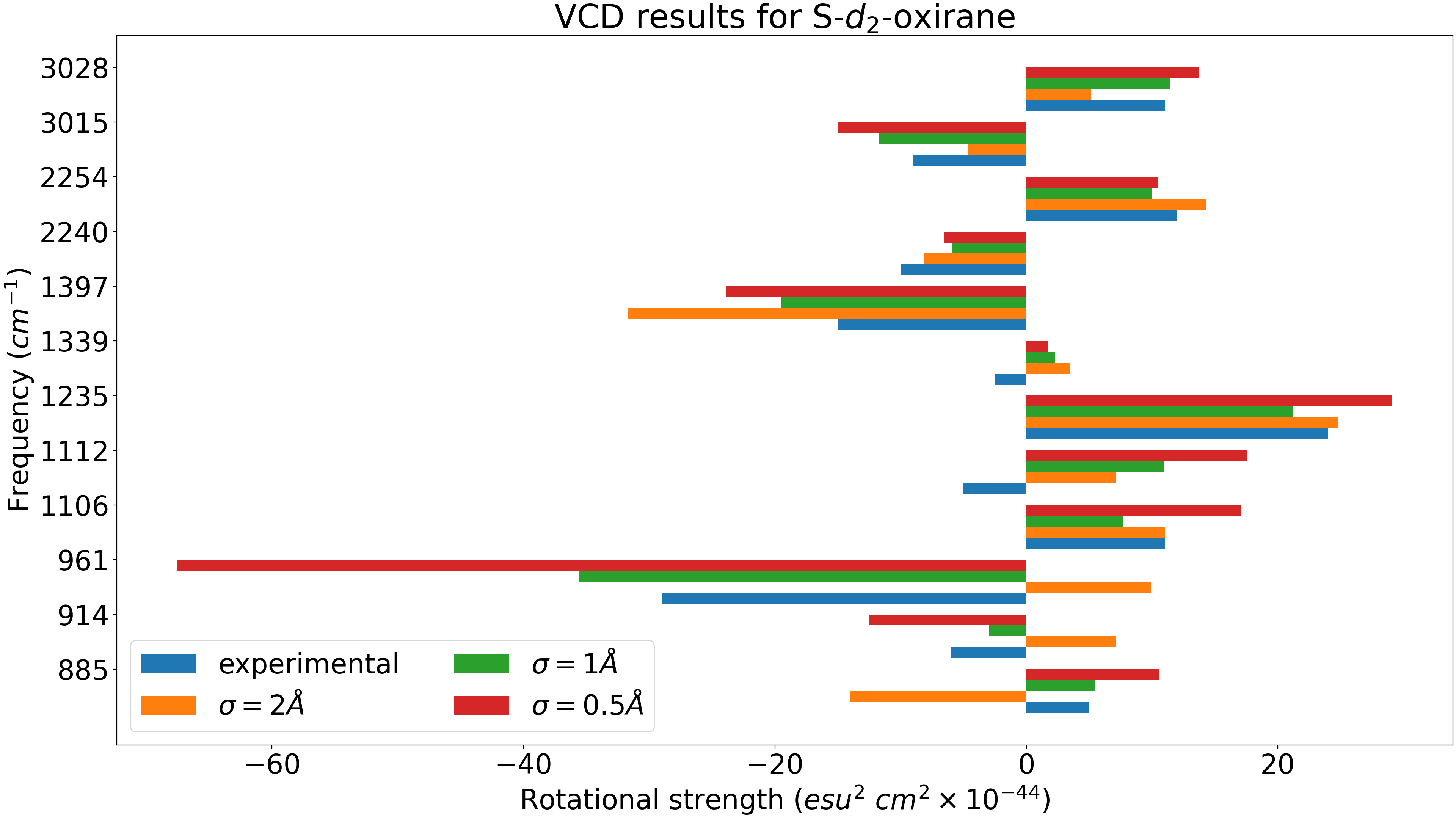}
    \caption{VCD results for the Hirshfeld partitioning with different values of $\sigma_A$ in Eq. \ref{hirshfeld_def}.
    Here we explore $\sigma =$ 0.5\AA, 1\AA \, or 2\AA.  
   The mean absolute error (relative to experiment) is 10.1 for $\sigma$= 2\AA, 4.22 for $\sigma$= 1\AA, and 9.26 for $\sigma$= 0.5\AA. Errors are in esu$^2$ cm$^2 \times 10^{-44}$. Note that one must be careful to choose $\sigma$ around 1 \AA \; for optimal results.}
    \label{VCD4}
\end{figure}

Inevitably, one cause for concern when performing PS calculations is how to choose the parameters for the $\left\{ \theta_A \right\}$ functions, e.g. the $\sigma_A$ parameter in Eq. \ref{hirshfeld_def}. This parameter dictates the amount of electron sharing between atoms: a smaller $\sigma_A$ parameter indicates a sharper cutoff between atoms. 
 To that end, 
 Figures \ref{VCD3} and \ref{VCD4} now show how choosing the cutoff between atoms affects VCD results for the Becke shifted and Hirshfeld partitionings, respectively. 
 One of the  strengths of the Becke scheme is that, whereas the functional form in Eq. \ref{hirshfeld_def} requires that one fix a continuous parameter $\sigma_A$,  for the Becke partitioning scheme, one must choose only an integer. Namely, one must only indicate the number of iterations of the function:
\begin{align}
    p(\nu) = \frac{3}{2}\nu-\frac{1}{2}\nu^3\label{becke_function}
\end{align}
More iterations makes a sharper cutoff. Typically, one uses three iterations\cite{becke:1988:jcp:beckegrid}. That being said, one expects VCD PS calculations based on Becke weights to be fairly robust because the partitioning is so robust.
Indeed,  according to the data in Fig. \ref{VCD3}, Becke partitioning is quite insensitive to the number of iterations. The only change in sign is the signal at 885 cm$^{-1}$ when 2 iterations are used. That being said, as shown in Fig. \ref{VCD4}, the data for Hirshfeld partitioning can be  sensitive to the choice of $\sigma_A$, which is a strong endorsement of the former over the latter.

One interesting observation from  the data in Table \ref{dz_table}) is that using the Bragg-Slater radii to shift the Becke weights significantly improves results when compared to using uniform weighting for all nuclei. This result is expected because the influence of each nucleus on an electron is not only a function of distance, but also the charge of the nucleus. Note that using Bragg-Slater radii has previously been found to yield accurate dipole moments; see Ref. \citenum{Mei:2015:partitioning_polarizablility}.

Finally, a word is now appropriate about the required size and resolution of the grid. In the Appendix, we  show how the VCD error depends on grid size. 
For the calculations here, if one seeks an error of 0.1\%, then one can use only 25 radial and 170 angular points. The data in Figs. \ref{VCD1}, \ref{VCD3}, and \ref{VCD4} (and Table \ref{dz_table}) use 100 radial and 590 angular points, which is very converged.
Second, as a matter of comparison, in the Appendix we have also plotted the relative error in the energy that one finds for an analogous DFT energy calculation.  Although the scales are different, note that the functional form is roughly the same. Using the same  grid size for the DFT calculation  as for $\bGamma$ above (namely 100 radial by 590 angular points), one would find a $1.04\times 10^{-9}$ absolute relative energy error in DFT.

Lastly, we consider timings. The total time to calculate $\partial {\bf m}/\partial \bP$ for the VCD calculations above  on a single processor was 1.8s for the cc-pvdz calculations. %and 18s for cc-pvtz. 
Thus, at the end of the day, whereas the data in Refs. \citenum{duston:2024:jctc_vcd} and \citenum{zhentao:2024:jcp:vcd_basis_free} required minutes or hours on multiple cores, PS is now quite competitive; at present, the cost for PS is slightly less than the cost for a standard Q-CHEM\cite{qchem6} MFP calculation. In fact, it would appear that PS dynamics over large system sizes will soon be possible.

\section{Discussion and Conclusions}

We have demonstrated that PS electronic structure theory can be naturally incorporated into modern electronic structure packages using the Becke weights for the partitioning functions and the standard DFT grids to evaluate matrix elements in an AO basis. Moreover, with the proper weights (and radii), we have secured just as much accuracy as before but now with a more robust scheme (less sensitive to parameterization).   Stability with basis set is as good as or slightly better than was found previously with Hirshfeld weight.  Perhaps most importantly, the total cost of a PS GHF calculation is roughly the same per iteration as the cost of a standard complex, GHF simulation.   Thus, the present paper should be of immediate relevance to the electronic structure community, especially those interested in electronic-nuclear processes.   We intend to release this code with the next major Q-Chem release. 

 Looking forward, many opportunities arise.
First, the calculations presented above were performed with Hartree-Fock. That being said, for many systems, one needs to invoke a higher level of theory that accounts for electron-electron correlation. Shumberger {\em et al} have already had some success extending MFP to second order Møller–Plesset\cite{Shumberger:2025:MP2_VCD} calculations, but the cost becomes  expensive and coupled cluster theory would be even more so. It will be  very worthwhile to investigate if phase space electronic structure methods can offer strong accuracy at a dramatically reduced cost for correlated methods.
 Second, 
 because phase space electronic structure theory introduces a small mass-dependent correction to the electronic Schrodinger equation, one would expect to find the most important corrections for systems with small electronic gaps, e.g. metallic solids. 
 Thus, solid state applications will be an important next step.  
Interestingly, Becke weights have been implemented recently for solids within an AO basis\cite{Wang:2024:periodic_gaussian_orbital}. 
Third, one of the strengths of PS is momentum conserving dynamics. For this purpose, the only necessary next step will be to evaluate gradients, $\frac{\partial \theta}{\partial R}$. Thus, the present developments should be of interest to those  who wish to run dynamics that conserve linear and angular momentum.

Fourth, magnetic fields represent an important area for future study. Magnetic field effects  are natural candidates for phase space electronic structure calculations because only moving charges feel magnetic fields and, within BO theory, electrons and nuclei are treated quite unequally such that,  after diagonalizing the electronic Hamiltonian, the electron can no longer experience a magnetic field when the nuclei start moving.
That being said, by parameterizing the electronic states around both $\bR$ and $\bP$, phase space electronic structure theory avoids this problem so that the electron can feel experience the direct, external applied magnetic field as well as the indirect magnetic field that arises from nuclear motion. Future work will necessarily need to incorporate Gauge-invariant atomic orbitals (GIAOs)\cite{pulay:2007:giao} to treat such magnetic field effects, but one can anticipate many future calculations with interesting magnetic field experiments, e.g. the Einstein-de Haas effect.

Finally, in this paper (and in all previous {\em ab initio} work in our groups), we have not considered the $\hbGamma \cdot \hbGamma/(2M)$  operator in Eq. \eqref{PS_ham}. In principle, this operator involves both one and two electrons terms (just like one finds for the self-energy of the polaritonic problem\cite{2025:Fabri:polariton_DSE}), so that the latter dress the standard electron-electron repulsion.  (Note that Ref. \citenum{xuezhi:cpr:review:2026} was wrong insofar as it concluded that one could treat the $\hbGamma \cdot \hbGamma/(2M)$ operator as a one-electron term). Thus, in the future, it will be very interesting to see how electron-electron interactions are altered interaction with the nuclei, and whether   any hint of attraction and superconductivity can be discerned.

\section{Acknowledgments}
This work was supported by the National Science Foundation under Grants No. CHE-2422858 (JES) and CHE-2102071 (YS),  and the AMOS program of the US Department of Energy, Office of Science, Basic Energy Sciences, Chemical Sciences, Biosciences and Geosciences (TJM).

\section{Author Declarations}
The authors have no conflicts to disclose.

\section{Appendix: Basis Set Dependence}

\begin{figure}[ht]
    \centering
    \includegraphics[width=6.5in]{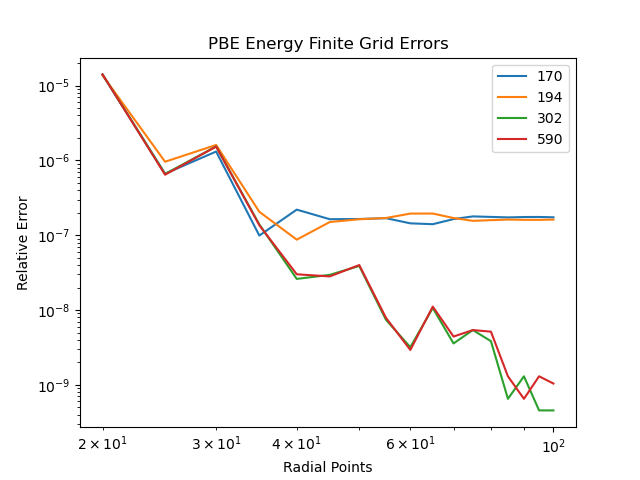}
    \caption{The error of the DFT ground state energy with the Perdew–Burke–Ernzerhof functional\cite{Perdew1996} as a function of Lebedev grid size for S-$d_2$-oxirane. The legend indicates the number of angular points. The error is computed as the absolute error between energies (compared to a reference with 500 radial and 2030 angular points) divided by the  absolute value of the energy of the reference. This figure is the only figure that uses a DFT functional in this paper. The basis is cc-pvdz.
    }
    \label{DFTErr}
\end{figure}

\begin{figure}[ht]
    \centering
    \includegraphics[width=6.5in]{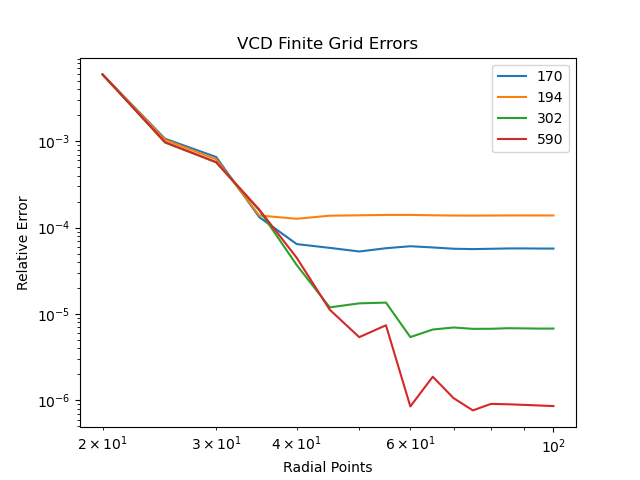}
    \caption{The error of VCD results as a function of Lebedev grid size for S-$d_2$-oxirane. The legend indicates the number of angular points. There are 15 VCD signals. The error is computed as the average absolute error between signals (compared to a reference with 500 radial and 2030 angular points) divided by the average absolute signal size of the reference. This measure is not the same as the average absolute relative error. While the VCD errors here  are larger than DFT energy errors in Fig. \ref{DFTErr}, the trends are similar and a huge grid is not required for a very reasonable VCD simulation. The calculation in this figure is from a HF ansatz with basis cc-pvdz. 
    }
    \label{VCDErr}
\end{figure}

In this Appendix, we now provide two figures. 
In Fig. \ref{DFTErr} we present the absolute relative error of the DFT energy (using the Perdew–Burke–Ernzerhof functional\cite{Perdew1996}) as a function of the number of grid points. This figure is presented for comparison to Fig. \ref{VCDErr} which shows the error of the VCD signal as a function of grid size. More specifically, for a PS calculation with Bragg-Slater shifted Becke weights, we plot the average absolute error between VCD signals (compared to a reference with 500 radial and 2030 angular points) divided by the average absolute signal size of the reference.   The absolute relative error is lower for the DFT energy versus the VCD signal, which is not surprising; for calculations that optimize energy, energy is always less sensitive than are other wavefunction properties. That being said,
both figures display qualitatively similar trends and one generate quite reasonable VCD spectra without enormous grids.


\bibliography{finalbib}
\end{document}